\documentclass[aps,pra,twocolumn]{revtex4-2}
\usepackage{times}
\usepackage{geometry} 		
\usepackage{graphicx}
\usepackage{amssymb,amsmath}
\usepackage{physics}
\usepackage{subfigure}
\usepackage{ulem}
\usepackage[dvipdfmx]{hyperref}
\usepackage[dvipdfmx]{color}
\usepackage{url}
\usepackage{braket}
\usepackage{framed}
\usepackage{enumerate}
\usepackage{autobreak}
\allowdisplaybreaks
\begin{document}
\title{Anonymous quantum sensing}
\author{Hiroto Kasai$^{1,2}$}
\email{kasai.quantum-loki-network@aist.go.jp}
\author{Yuki Takeuchi$^{3}$}
\author{Hideaki Hakoshima$^{2}$}
\author{Yuichiro Matsuzaki$^{2}$}
\email{matsuzaki.yuichiro@aist.go.jp}
\author{Yasuhiro Tokura$^{1}$}
\email{tokura.yasuhiro.ft@u.tsukuba.ac.jp}
\affiliation{
$^{1}$
Graduate School of Pure and Applied Sciences, University of Tsukuba, 1-1-1 Tennodai, Tsukuba, Ibaraki 305-8571, Japan
}
\affiliation{
$^{2}$
Research Center for Emerging Computing Technologies, National institute of Advanced Industrial Science and Technology (AIST), Central2, 1-1-1 Umezono, Tsukuba, Ibaraki 305-8568, Japan
}
\affiliation{
$^{3}$
NTT Communication Science Laboratories,
NTT Corporation, 3-1 Morinosato Wakamiya, Atsugi, 
Kanagawa 243-0198, Japan
}
\begin{abstract}
A lot of attention has been paid to a quantum-sensing network for detecting magnetic fields in different positions. 
Recently, cryptographic quantum metrology was investigated where the information of the magnetic fields is transmitted in a secure way. 
However, sometimes, the positions where non-zero magnetic fields are generated could carry important information. 
Here, 
we propose an anonymous quantum sensor where an information of positions 
having non-zero magnetic fields is hidden after measuring magnetic fields with a quantum-sensing network. 
Suppose that agents are located in different positions and they have quantum sensors. 
After the quantum sensors are entangled, 
the agents implement quantum sensing that provides a phase information if non-zero magnetic fields exist, 
and POVM measurement is performed on quantum sensors.
Importantly, 
even if the outcomes of the POVM measurement is stolen by an eavesdropper, 
information of the positions with non-zero magnetic fields is still unknown for the eavesdropper in our protocol.
In addition, we evaluate the sensitivity of our proposed quantum sensors 
by using Fisher information when there are at most two positions having non-zero magnetic fields.
We show that the sensitivity is finite 
unless these two (non-zero) magnetic fields have exactly the same amplitude.
Our results pave the way for new applications of quantum-sensing network.
\end{abstract}
\maketitle
\section{Introduction}
Protecting privacy information is an important subject.
Not only the contents of the messages but also information about who transmits the message could be considered as private information.
The former is protected by encryption and the latter is protected by anonymization. 
About the latter,
we can enumerate Tor (The Onion Routing) 
 \cite{dingledine2004tor} 
 as a famous classical anonymous communication technology.
Tor protocol supports anonymization with public key encryption,
and so a quantum computer can break the public key and deanonymize Tor users.
Then, 
we need quantum anonymous communication technology which has information-theoretic security. 
Such quantum anonymous communication schemes with GHZ state 
\cite{christandl2005quantum} 
or W state 
\cite{lipinska2018anonymous} 
have been proposed. 
There are other anonymous communication technologies using other quantum states 
\cite{
newbrassard2007anonymous,
unnikrishnan2019anonymity,
wang2010economical}
.
These protocols allow only one sender to send a message anonymously at the same time.\par
Theoretical and experimental researches about quantum sensing with qubits have been carried out \cite{
maze2008nanoscale,
balasubramanian2008nanoscale,
taylor2008high,
kominis2003subfemtotesla,
bal2012ultrasensitive,
eldredge2018optimal,
proctor2018multiparameter}.
Due to the excellent sensitivity and spatial resolution,
the quantum sensors are expected to be used in medical science and material engineering \cite{
mitchell2020colloquium,
schirhagl2014nitrogen,
barry2016optical} .
Especially, quantum magnetic-field sensing could be applied 
to magnetoencephalography \cite{xia2006magnetoencephalography},
and so we might collect private information with quantum sensors in the future.
This means that security about the information obtained by quantum sensors could be important.
Because of this reason, 
recent research has been devoted to hybridize quantum sensing and quantum communication 
to add security in sensing tasks 
\cite{
degen2017quantum,
giovannetti2001quantum,
giovannetti2002quantum,
giovannetti2002positioning,
chiribella2005optimal,
chiribella2007secret,
huang2019cryptographic,
xie2018high,
shettell2021cryptographic,
giovannetti2006quantum,
giovannetti2011advances,takeuchi2019quantum,
pirandola2020advances,
okane2020quantum,
yin2020experimental}
.
There are also researches to add novel functions (such as estimating the spatial distribution of magnetic fields)
 with quantum sensing by using quantum networks \cite{
zhao2020field,
brida2010experimental,
perez2012fundamental,
baumgratz2016quantum,
komar2014quantum}.\par
We propose ``anonymous quantum sensor'' that is a
new protocol to combine quantum anonymous communication and quantum sensor.
Anonymous quantum sensor gives functionality that 
information of positions having non-zero magnetic fields is hidden by
using entanglement among quantum sensors in multiple positions.
In the context of conventional research of quantum anonymous communication,
an advantage of our protocol is that two senders can send anonymously continuous quantities at the same time.
Concretely, this protocol is carried out as below.
First, there are non-zero magnetic fields in one or two positions 
and 
no magnetic fields in the other positions out of $n$ positions.
Second, 
distribute entangled qubits in each positions out of $n$ positions
and
entangle each quantum sensor of each participant.
Then, we assume that the protocol uses Dicke states \cite{
hepp1973superradiant,
dicke1954coherence,
bartschi2019deterministic}.
Third, 
each participant writes information on magnetic fields 
on phase information of the entangled state
by interacting own quantum sensors with magnetic fields in their positions.
Finally, 
a specified POVM measurement of the entangled state gives information of magnetic fields.
An important point is 
that the probability of measurement outcome is symmetric with 
respect to each participant with property of entanglement.
Thus, 
even if the outcomes of the POVM are broadcasted (or are stolen by eavesdropper), 
information of the positions with non-zero magnetic fields 
is still unknown for the eavesdropper in our protocol.
Then, 
if the actual number of the positions with non-zero magnetic fields is
more than what we assumed 
(for example, there are three positions with non-zero magnetic fields),  
we cannot estimate correct values of magnetic fields but confidentiality 
of the positions with non-zero magnetic fields is kept.
\section{
Magnetic-field sensing with a qubit
}
Let us review how to measure the amplitude of the magnetic fields with qubits.
\subsection{
The dynamics of a qubit under the magnetic fields
}
The dynamics of the qubit to interact with the magnetic fields is described as follows.
Let us define $\sigma_{z}$ and $\sigma_{x}$ be the standard Pauli operators.
The Hamiltonian of the qubit is represented as
\begin{eqnarray}
\hat{H}
&=&
\frac{\omega}{2} \sigma_{z}.
\end{eqnarray}
We assume that a shift of the resonant frequency $\omega$ is proportional to the applied magnetic fields.
We choose $ \rho_{0} = \ket{ + }\bra{ + } $ , 
an eigenstate of $\sigma_{x}$ , as the initial state.
The state evolves by the unitary operator of $\hat{U} = \exp (-i\hat{H} t)$. 
It is worth mentioning that $t$ denotes the interaction time with the magnetic fields,
and so $t$ is a known parameter. 
Throughout this paper, 
we assume $\hbar=1$ and notation that double sign applies in the same order.
After the time evolution, 
we obtain the quantum state as follows.
\begin{eqnarray}
{}
&{}&
\rho_{0,\omega}
\nonumber \\
&=&
\hat{U}
\rho_{0}
\hat{U}^{\dagger}
\nonumber \\
&=&
\biggl\{
 \frac{1}{\sqrt{2}} \bigl( \ket{0} + e^{ i\omega t } \ket{1} \bigr)
\biggr\}
\biggl\{ 
 \frac{1}{\sqrt{2}} \bigl( \bra{0} + e^{ -i\omega t } \bra{1} \bigr)
\biggr\}
\nonumber \\
\end{eqnarray}
The relative phase is encoded in the quantum state, 
and the phase contains the information of the resonant frequency (corresponding to the applied magnetic fields).
\subsection{
Evaluation of estimation uncertainty
}
We explain how to readout the relative phase of the quantum state,
and also introduce a way to quantify the estimation uncertainty of the resonant frequency 
by using classical Fisher information.
The definition of the classical Fisher information and details 
to calculate the estimation uncertainty 
are described in Appendix \ref{sec:fisher-info_and_Cramer-Rao}. 
${}$\par
As we explained, 
the quantum state after the time evolution is described as following.
\begin{eqnarray}
\rho_{0,\omega}
&\equiv&
\hat{U}
\rho_{0}
\hat{U}^{\dagger}
\end{eqnarray}

The unknown parameter to be estimated by the protocol is as follows.
\begin{eqnarray}
\theta
&=&
\omega t
\end{eqnarray}
We choose the following POVM measurement for the readout.
\begin{eqnarray}
\begin{cases}
\hat{E}_{1}
=
 \ket{ + }\bra{ + } 
\\
\hat{E}_{2}
=
\hat{I} - \hat{E}_{1} 
\end{cases}
\end{eqnarray}
Since we have $
\braket{ + | \hat{U} | + }
=
\cos { 
 \frac{ \theta }{2}
 }
 $,
the probability is described as follows.
\begin{eqnarray}
\begin{cases}
P_{1}
=
\cos^{2} { \frac{ \theta }{2} }
\\
P_{2}
=
\sin^{2} { \frac{ \theta }{2} }
\end{cases}
\end{eqnarray}
Then, we calculate the matrix of classical Fisher information $J$.
\begin{eqnarray}
\begin{cases}
\frac{ \partial P_{1} } { \partial \theta } 
=
- \sin{ \frac{ \theta }{2} } \cos{ \frac{ \theta }{2} }
\\
\frac{ \partial P_{2} } { \partial \theta } 
=
 \sin{ \frac{ \theta }{2} } \cos{ \frac{ \theta }{2} }
\end{cases}
\end{eqnarray}
From the above calculation, we obtain $ J_{1,1} = 1 $.
The inverse matrix of classical Fisher information provides the uncertainty of the estimation,
and we obtain $ (J^{-1})_{1,1} = 1 $.
Then, Cram\'{e}r-Rao inequality is described as follows.
\begin{eqnarray}
(\Delta \theta)^{2}
&\ge&
\frac{1}{N}
\end{eqnarray}
$(\Delta \theta)^{2}$ denotes the variance, which we call an uncertainty of the estimation.
Here, 
we define that the number of the measurements is $N$ and
the sensitivity as the inverse of the estimation uncertainty.
\section{anonymous quantum sensor}
\subsection{Definitions and Model}
\label{sec:Definitions_and_Model}
In our protocol, 
we consider a distributer, a measurer, and participants.
Let us define these three.
\begin{enumerate}
\item Definition of distributer\\
A distributer is an entity that distributes entangled qubits 
to all participants which we explain below. 
Distribute one qubit per one participant.
\item Participants\\
A participant is defined as an entity that can be a sender which we define later.
Every participant is located in spatially different positions.
There are $n$ participants in this protocol, 
and the set of the participants is defined as $V \equiv \{1,2,\ldots, n\}$.
Each participant receives a qubit from the distributer, 
and the qubit interacts with the magnetic fields if non-zero
magnetic fields are generated at the position.\par
Senders are defined as some of participants that use the quantum sensor (qubit)
 in their position where non-zero magnetic fields are generated.
The information to be sent is the amplitude of the magnetic fields interacted with the qubit.\par
We consider a way for the senders to send the amplitude of 
the magnetic fields $\omega_{j}$ $(j= 1,2,\ldots, m)$ in an anonymous way 
that the information about which positions have non-zero magnetic fields is not revealed.
Here, 
$\omega_{j}$ denotes magnetic fields at the $j$-th position and $m$ denotes the number of positions 
that generate non-zero magnetic fields. 
Also, without loss of generality, 
we can assume the following relationship
\begin{eqnarray} 
0 < \omega_{1} \le \omega_{2} \le \ldots \le \omega_{m} .
\end{eqnarray} 
Then, we define a set of senders $S ( \subset V )$ as follows
\begin{eqnarray} 
S \equiv \{ s_{1},s_{2},\ldots, s_{m} \}
\end{eqnarray} 
where we have $m=|S|$.
This means that the $j$-th sender will send the information of $\omega_j$ for $j=1,2,\ldots,m$.\par
A non-sender is defined as a participant located at a position with zero magnetic fields.
A set of non-senders is defined as $R \equiv \bar{S} ( = V \setminus S)$.
\item Definition of measurer\\
A measurer does not belong to the set of the participants $V$.
The measurer will perform a POVM measurement on the qubits of all participants.
Also, 
the measurer performs an estimation 
(by using a method such as a maximum likelihood estimation method) 
about the magnetic fields based on the outcomes of the POVM measurements.
\end{enumerate}
\subsection{Protocol flow}
\begin{figure}[htbp]
 \includegraphics[height=9.0cm,width=6.2cm]
{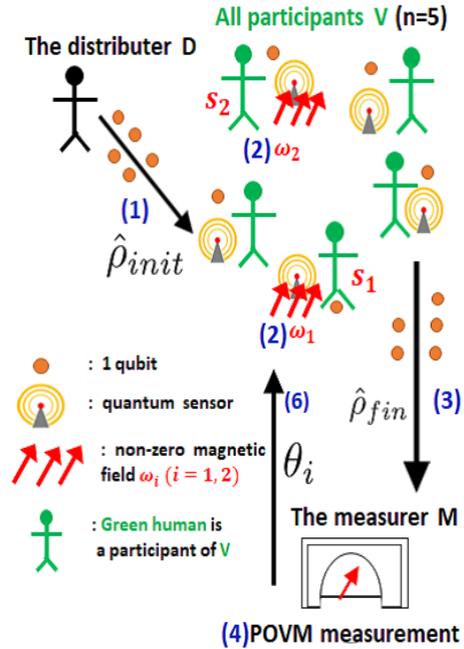}
 \caption{
 Schematic of our anonymous quantum sensor.
 }
 \label{pic:protocol-flow}
\end{figure}
We illustrate our scheme in the figure \ref{pic:protocol-flow}.
We assume that each participant is located in a different position,
and the participant operates a qubit 
as a magnetic-field sensor at his/her position.\par
Throughout this paper,
we assume that there is no decoherence and the quantum operations are perfect.
\begin{oframed}
Goal: 
each sender $s_{j}$ broadcasts the information of $\omega_{j}$ to all participants $V$ in an anonymous way.
\begin{enumerate}
\item[(1)] The distributer $D$ distributes entangled qubits to all participants $V$, and each participant receives one of the qubits.
Each participant interacts his/her qubit
with local magnetic field in each position.
\item[(2)] Each participant sends his/her qubit to the measurer $M$.
\item[(3)] The measurer $M$ 
implements POVM on all qubits sent from the participants.
\item[(4)] 
The steps from 1 to 4 are repeated, 
and the measurer $M$ obtains the distribution of all measurement outcomes.
\item[(5)]The measurer $M$ estimates 
the
values of unknown parameters $\theta_{i} \quad (i=1,2,\ldots, m)$
from the distribution of all measurement outcomes.
\item[(6)] The measurer $M$ broadcasts 
the
estimated values of $\theta_{i} \quad (i=1,2,\ldots, m)$ to all participants $V$ via classical channel.
\item[(7)] 
Every participant knows the estimated values 
of $\omega_{i} \quad (i=1,2,\ldots, m)$ 
from the broadcasted information 
of $\theta_{i} \quad (i=1,2,\ldots, m)$.
\end{enumerate}
\end{oframed}
\subsection{Required security condition}
The measurer $M$ obtains 
classical information by implementing the POVM. 
However, the classical information can be copied in anywhere including unknown environment
 that cannot be controlled. 
If an eavesdropper (Eve) tries to steal the information, 
Eve can access such an environment,
 which is not easy to prevent.
It is desirable to protect the information about which positions generate the magnetic fields
 even when the classical information is stolen by Eve. 
Therefore, 
we adopt the following definition as a condition of security in our paper.
Tracelessness is defined as follows. 
Even if Eve steals all classical information that a distributer, 
participants, 
and a measurer obtain during the protocol, 
the information about which positions generate magnetic fields is unknown for Eve. 
In our protocol, 
only a measurer obtains the classical information. 
So the tracelessness is guaranteed if the results of the POVMs are independent 
of the information about which positions generate magnetic fields. 
When taraceless is guaranteed, such a protocol is typically considered as annoymous. 
Finally, 
we assume that algorithm of our protocol is known to Eve.
\subsection{Set up of protocol}
We consider that there are $n$ participants and $m$ senders 
(we assumed the condition: $ m \le \bigl[ \frac{n+1}{2} \bigr] $
and $[\cdot]$ is floor function.
), and
\begin{eqnarray}
\rho_{init}
&=&
\sum_{i=0}^{ \bigl[ \frac{n}{2} \bigr] }
 q_{i} 
 \ket{ \phi^{ n }_{ i ,+} } \bra{ \phi^{ n }_{ i ,+} }
\end{eqnarray}
where
\begin{eqnarray}
\begin{cases}
0 \le q_{i} \quad (i = 0,1,\ldots , \bigl[ \frac{n}{2} \bigr] )
\\
\sum_{i=0}^{ \bigl[ \frac{n}{2} \bigr] } q_{i} = 1
\end{cases}
\end{eqnarray}
$\ket{ \phi^{n}_{k,l} }$ is a superposition of Dicke states 
(
see Appendix \ref{sec:dicke_state}
and 
\ref{sec:anonymity-of-probability}
).\par
The qubits of the participants in the set of senders $S$ evolve 
by the following Hamiltonian
\begin{eqnarray}
\label{eq:all_hamiltonian}
\hat{H}
&=&
\sum_{j \in S} \hat{H_{j}}
=
\frac{1}{2} \sum_{j =1}^{m} \omega_{j} \hat{\sigma}_{z,s_{j}}
\end{eqnarray} 
where $ \omega_{j} 
(> 0 )
\quad 
(j=1,2,\ldots,m ) $ 
denotes magnetic fields, 
already defined in Section \ref{sec:Definitions_and_Model}.
Then, 
the unitary operator is represented as follows.
\begin{eqnarray}
\label{eq:all_time_evolution}
\hat{U}
&=&
e^{ -i\hat{H}t }
=
\prod_{j =1}^{m} 
e^{ -\frac{1}{2} i\omega_{j} t \hat{\sigma}_{z,s_{j} } }
\end{eqnarray} 
This unitary operator contains only $\hat{I}$ and $\hat{\sigma}_{z}$ without bit-flip operators 
such as $\hat{\sigma}_{x}$. 
The quantum state after the evolution is described as follows.
\begin{eqnarray}
\rho_{fin}
&=&
\hat{U}
\rho_{init}
\hat{U}^{\dagger}
\end{eqnarray}
We consider a POVM measurement on the state explained above.
We choose the POVM as follows.
\begin{eqnarray}
\begin{cases}
\hat{E}_{i,\pm}
=
c_{i,\pm}
 \ket{ \phi^{ n }_{ i ,\pm} } \bra{ \phi^{ n }_{ i ,\pm} }
 \quad ( i=0,1,\ldots, \bigl[ \frac{n}{2} \bigr] )
\\
\hat{E}_{f}
=
\hat{I}
-
\sum_{i =0}^{ \bigl[ \frac{n}{2} \bigr] } 
\sum_{j=\pm} 
\hat{E}_{i,j}
\end{cases}
\nonumber \\
\end{eqnarray}
where $c_{i,j} = 0,1 (i=0,1,\ldots, \bigl[ \frac{n}{2} \bigr] , j=\pm)$
is satisfied. 
We can calculate the probability of the POVM measurement as follows.
\begin{eqnarray}
\label{eq:prob_povm}
\begin{cases}
P_{i , \pm }
=
\mathrm{Tr}
(
\hat{E}_{i , \pm} 
\rho_{fin}
)
=
\pm c_{i,\pm} q_{i}
( \gamma^{n,m}_{ i , \pm } )^{2}
 \\
 \quad
 (i=0,1,\ldots, \bigl[ \frac{n}{2} \bigr] )
 \\
P_{f}
=
1
-
\sum_{i =0}^{ \bigl[ \frac{n}{2} \bigr] }
\sum_{j=\pm}
P_{i,j}
\end{cases}
\end{eqnarray}
By choosing suitable coefficients of $c_{i,j}$ (0 or 1), 
we can control whether the form of
$( \gamma^{n,m}_{ i , \pm } )^{2} $ 
appears or not in the probability distribution 
 ( $\gamma_{i,\pm}^{n,m}$ is explained in Appendix \ref{sec:anonymity-of-probability}. )
. 
So, 
 our purpose is to obtain a desired probability distribution by changing the values of $c_{i,j}$.
 Then, 
 we calculate the matrix of classical Fisher information and inverse matrix of that 
 from the probability distribution.
This allows us to evaluate the estimation uncertainty of unknown parameters. 
Throughout of our paper, 
 we call $c_{i,j}$ ``POVM parameter".
For any index $i ( \in \{ 0,1,\ldots, \bigl[ \frac{n}{2} \bigr] \} ) $ satisfying $c_{i,j}=0$, the terms $ q_{i} \ket{ \phi^{ n }_{ i ,+} } \bra{ \phi^{ n }_{ i ,+} }$ in the initial state do not contribute probability.
So, we assume $q_{i} =0 $.
\begin{eqnarray}
{}
&{}&
c_{i,j}=0
\quad 
 ( \forall i \in F \subset \{ 0,1,\ldots, \bigl[ \frac{n}{2} \bigr] \} )
 \nonumber \\
&\Rightarrow&
 q_{i} = 0 
 \quad 
 ( \forall i \in F )
\end{eqnarray}
The number of the unknown parameters to be estimated is $m$,
 which is also the number of positions with non-zero magnetic fields. 
For a given $m$,
we need to choose suitable POVM parameters $c_{i,j}$.
\subsection{Analytics}
\subsubsection{1 position with non-zero magnetic field ($m=1$) }
First, we consider a case of $m=1$. 
We can interprete this as an anonymous transmission of one continuous value. 
Then, the initial state is as follows.
\begin{eqnarray}
\rho_{0}
&=&
\ket{ \phi^{n}_{0,+} } \bra{ \phi^{n}_{0,+} }
\end{eqnarray}
where
$
\ket{ \phi^{n}_{0,+} } 
=
\frac{1}{\sqrt{2}}
( \ket{0^{n}} + \ket{1^{n}} )
$
denotes the GHZ state. 
We consider a case that a participant $j (\in V)$ is the sender, and the Hamiltonian is described as follows.
\begin{eqnarray}
\hat{H}
&=&
 \frac{w}{2} \hat{\sigma}_{z,j}
\end{eqnarray}
The unitary operator is described as follows.
\begin{eqnarray}
\hat{U}
&=&
e^{-
\frac{it }{2} w \hat{\sigma}_{z,j}
}
\end{eqnarray}
After the time evolution, 
the quantum state is desdribed as follows.
\begin{eqnarray}
\rho_{0,\omega}
&=&
\hat{U}
\rho_{0}
\hat{U}^{\dagger}
\end{eqnarray}
We consider to implement POVMs on this state.
The parameter to be estimated is as follows.
\begin{eqnarray}
\theta
&=&
\omega t
\end{eqnarray}
Then, we choose the following POVM elements.
\begin{eqnarray}
\begin{cases}
\hat{E}_{1}
=
 \ket{ \phi^{n}_{0,+} }\bra{ \phi^{n}_{0,+} } 
\\
\hat{E}_{2}
=
\hat{I} - \hat{E}_{1} 
\end{cases}
\end{eqnarray}
We obtain the following equation.
\begin{eqnarray}
\braket{ \phi^{n}_{0,+} | \hat{U} | \phi^{n}_{0,+} }
&=&
\cos { 
 \frac{ \theta }{2}
 }
\end{eqnarray}
Probability of each POVM element is calculated as follows.
\begin{eqnarray}
\begin{cases}
P_{1}
=
\cos^{2} { \frac{ \theta }{2} }
\\
P_{2}
=
\sin^{2} { \frac{ \theta }{2} }
\end{cases}
\end{eqnarray}
as explained in Sec. \ref{sec:evaluation-of-uncertainty}.
Cram\'{e}r-Rao inequality is obtained as follows.
\begin{eqnarray}
(\Delta \theta)^{2}
&\ge&
\frac{1}{N}
\end{eqnarray}
In this case, 
the quantum state after the interaction with the magnetic fields is independent 
of $j$ (an identity of the sender). 
So any POVM element cannot reveal the information about which positions generate the magnetic fields. 
This means that, for $m=1$, 
even if Eve performs state tomography on the quantum state 
after the interaction with the magnetic fields, 
the tracelessness is still guaranteed.\par
The above discussion is equal to the case where POVM parameters are chosen as follows
\begin{eqnarray}
 \begin{cases}
 c_{0,+} = 1 
 \\
 c_{i,j} = 0 \quad ( \text{otherwise} )
\end{cases}
\end{eqnarray}
where the normalization condition is chosen as follows
\begin{eqnarray}
q_{0} 
=
1.
\end{eqnarray}
\subsubsection{
2 positions with non-zero magnetic fields ($m=2$)
}
Secondly, 
we consider a case of $m=2$,
where the number of the positions with non-zero magnetic fields is two.
Equivalently, we have two senders. 
For $|S|=2$,
we rewrite the unknown parameter as follows.
\begin{eqnarray}
\begin{cases}
\theta_{1}
\equiv
(
\omega_{1} + \omega_{2}
)
t
\\
\theta_{2}
\equiv
(
\omega_{1} - \omega_{2}
)
t
\end{cases}
\end{eqnarray}
Since we will consider a case where there are many participants,
we assume $5 \le n$, and this is equivalent to $|R|>|S|=2$.
\begin{eqnarray}
 \begin{cases}
 c_{0,\pm} = c_{a,+} = 1 
 \quad
 ( |S| = 2 \le a \le \bigl[ \frac {n}{2} \bigr] )
\\
 c_{i,j} = 0 \quad ( \text{otherwise} )
\end{cases}
\end{eqnarray}
where the normalization condition is chosen as follows.
\begin{eqnarray}
q_{0} + q_{a}
=
1
\end{eqnarray}
The integer parameter $a$ should be chosen to optimize the sensitivity 
of this protocol.
Diagonal elements of the inverse matrix 
of classical Fisher information are obtained as follows.
\begin{eqnarray}
(J^{-1})_{1,1}
=
\frac{1}{q_{0}}
\end{eqnarray}
\begin{eqnarray}
{}
&{}&
(J^{-1})_{2,2}
\nonumber \\
&=&
\frac{1}
{
4 a^{2}(n-a)^{2}(1-q_{0})q_{0}
\sin^{2}{\frac{\theta_{2}}{2}}
}
\biggl[
(2a^{2}-2an -n 
\nonumber \\
&+&
 n^{2} )^{2}
 \sin^{2}{\frac{\theta_{1}}{2}}
+
4 q_{0}a(n-a)
\biggl\{
a(n-a) 
 \sin^{2}{\frac{\theta_{2}}{2}}
\nonumber \\
&+&
 ( 2a^{2}-2an-n+n^{2} )
(1 - \cos{\frac{\theta_{1}}{2}} \cos{\frac{\theta_{2}}{2}} )
\biggr\}
\biggr]
\nonumber \\
\end{eqnarray}
Suppose that $a$ is a function of $n$ ($a=a(n)$).
We define $\beta_{a,n}$ as follows. 
(see the
Appendix \ref{sec:evaluation-of-uncertainty} for the details).
\begin{eqnarray}
\beta_{a,n}
\equiv
\frac{a}{n}
\end{eqnarray}
Next, 
suppose that $\beta_{a,n}$ and $n$ are the tunable parameters.
Then,
we rewrite $(J^{-1})_{2,2}$ to $(J^{-1})_{2,2}^{n , \beta}$ as follows.
\begin{eqnarray}
{}
&{}&
(J^{-1})_{2,2}^{n,\beta}
\nonumber \\
&=&
\frac{ 1 }
{ (1-q_{0} )
\sin^{2} { \frac{ \theta_{2} }{2} }
}
 \biggl[
\sin^{2} { \frac{ \theta_{2} }{2} } 
 \nonumber \\
&+& 
2 \biggl\{
\frac
{ 1- n^{-1} }
{ 2 \beta_{a,n} ( 1- \beta_{a,n} )}
- 1 \biggr\}
\biggl(
 1 - \cos { \frac{ \theta_{1} }{2} } \cos { \frac{ \theta_{2} }{2} }
 \biggr)
 \nonumber \\
&+& 
\biggl\{
\frac
{ 1- n^{-1} }
{ 2 \beta_{a,n} ( 1- \beta_{a,n} )}
 - 1
 \biggr\}^{2}
\frac{1}{ q_{0}}
\sin^{2} { \frac{ \theta_{1} }{2} }
 \biggr]
\nonumber \\
\end{eqnarray}
In order to minimize $ (J^{-1})_{2,2}^{n,\beta} $, 
we can choose $\beta_{a,n} = \frac{1}{n} \bigl[ \frac{n}{2} \bigr] $
(See Appendix \ref{sec:evaluation-of-uncertainty} for the details), 
and we obtain
\begin{eqnarray}
{}
&{}&
 (J^{-1})_{2,2}^{n, \frac{1}{n} \bigl[ \frac{n}{2} \bigr] }
\nonumber \\
&=&
\frac{ 1 }
{ (1-q_{0} )
\sin^{2} { \frac{ \theta_{2} }{2} }
}
 \biggl\{
\frac{ 1}{ q_{0}}
\biggl( 1- \frac{1} { \bigl[ \frac{n+1}{2} \bigr] } \biggr)^{2} 
\sin^{2} { \frac{ \theta_{1} }{2} }
\nonumber \\
&+&
2 \biggl( 1- \frac{1} { \bigl[ \frac{n+1}{2} \bigr] } \biggr)
\biggl(
1- \cos { \frac{ \theta_{1} }{2} } \cos { \frac{ \theta_{2} }{2} }
\biggr)
+
 \sin^{2} { \frac{ \theta_{2} }{2} } 
 \biggr\}
 \nonumber \\
\end{eqnarray}
In the limit of a large $n$, we have 
${\rm{lim}}_{n\rightarrow \infty} (J^{-1})_{2,2}^{n, \frac{1}{n} \bigl[ \frac{n}{2} \bigr] }=(J^{-1})_{2,2}^{\infty,\frac{1}{2} }$.
 So, we obtain the following.
\begin{eqnarray}
{}
&{}&
(J^{-1})_{2,2}^{\infty,\frac{1}{2} }
=
 \lim_{n \to \infty}
(J^{-1})_{2,2}^{n, \frac{1}{n} \bigl[ \frac{n}{2} \bigr] }
\nonumber \\
&=&
\frac{
 \sin^{2}{\frac{\theta_{1}}{2}}
+
q_{0}
(
2
-
2 \cos{\frac{\theta_{1}}{2}}\cos{\frac{\theta_{2}}{2}}
+
 \sin^{2}{\frac{\theta_{2}}{2}}
)
}
{
 (1-q_{0})q_{0}
\sin^{2}{\frac{\theta_{2}}{2}}
}
\nonumber \\
\end{eqnarray}
Then, we plot
$ \log_{10} (J^{-1})_{2,2}^{n,\frac{1}{2} } $ 
against ($ \theta_{1}, \theta_{2}$) and $n$ where we assume $n$ is an even number.
\begin{figure}[htbp]
 \includegraphics[height=6.4cm,width=7.4cm]
 {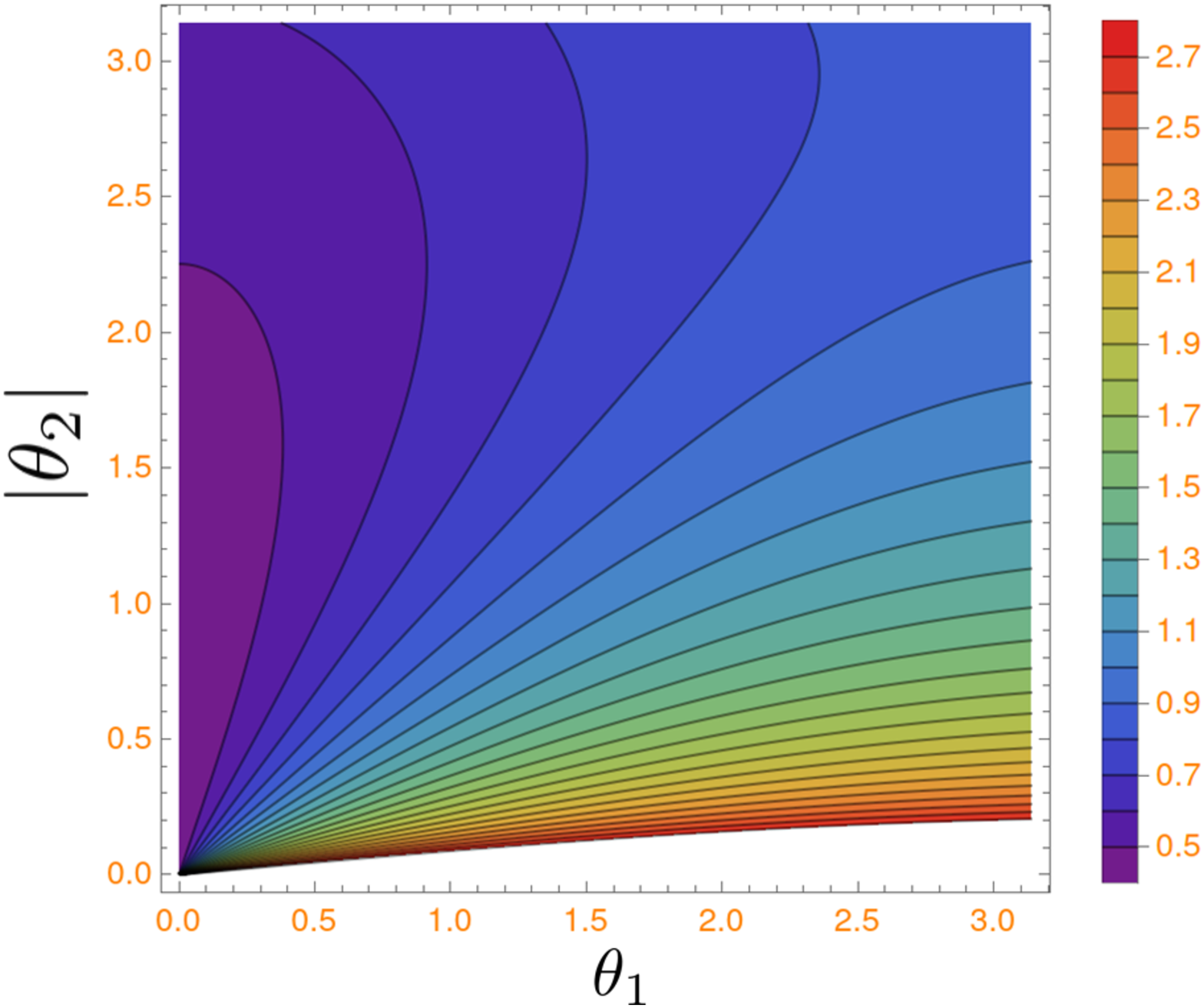}
 \caption{
Plot of $ \log_{10} (J^{-1})_{2,2}^{10^1,\frac{1}{2} } $ 
against $\theta_{1}$ and $\theta_{2}$ with a parameter of $q_{0}=0.33$.
}
 \label{fig:J_n=10^1,q_0=033}
 \end{figure}
\begin{figure}[htbp]
 \includegraphics[height=6.4cm,width=7.4cm]
 {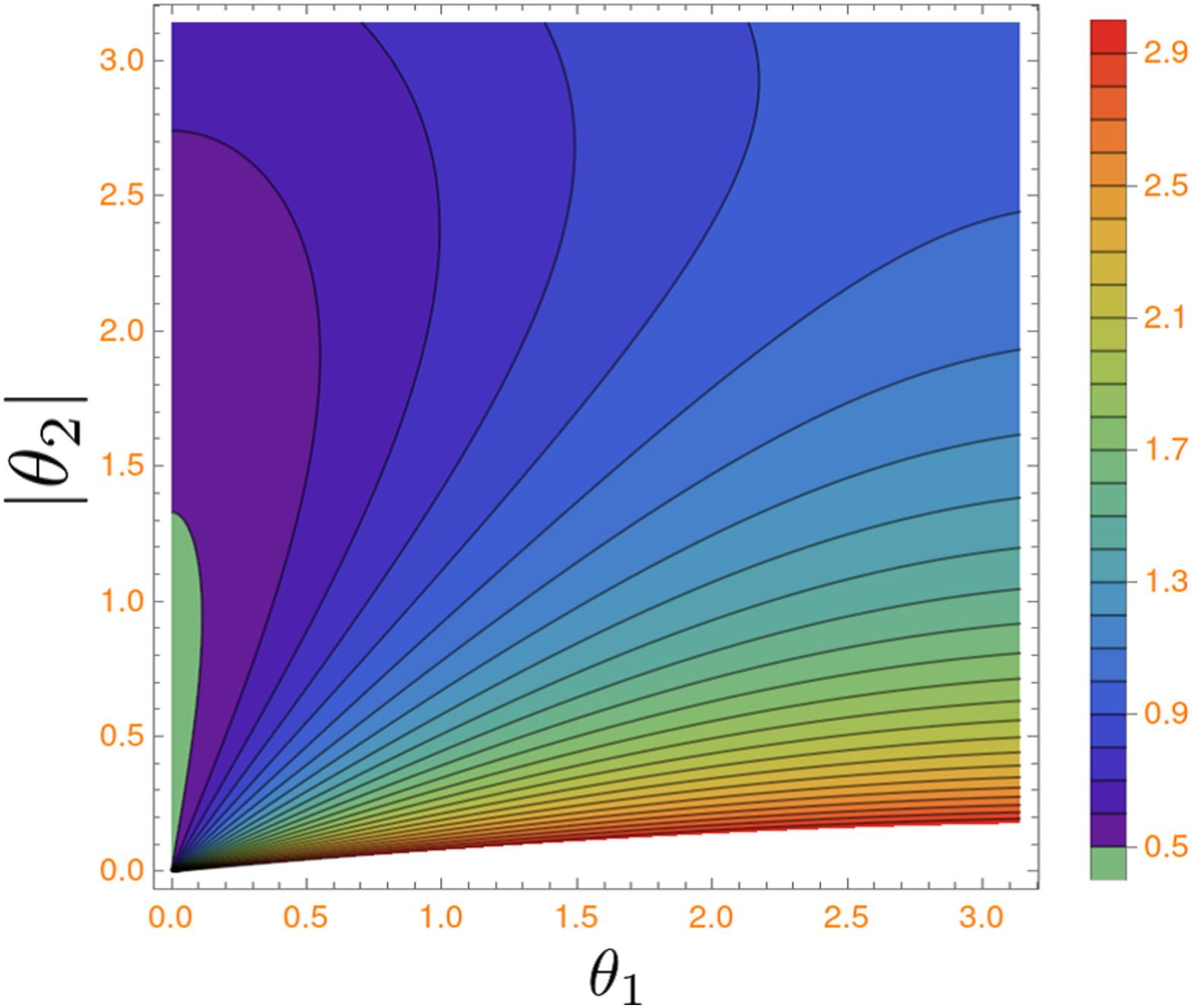}
 \caption{
Plot of $ \log_{10} (J^{-1})_{2,2}^{10^4,\frac{1}{2} } $ against $\theta_{1}$ and $\theta_{2}$ with a parameter of $q_{0}=0.33$. 
 }
 \label{fig:J_n=10^4,q_0=033}
 \end{figure}
\begin{figure}[htbp]
 \includegraphics[height=6.4cm,width=7.4cm]
 {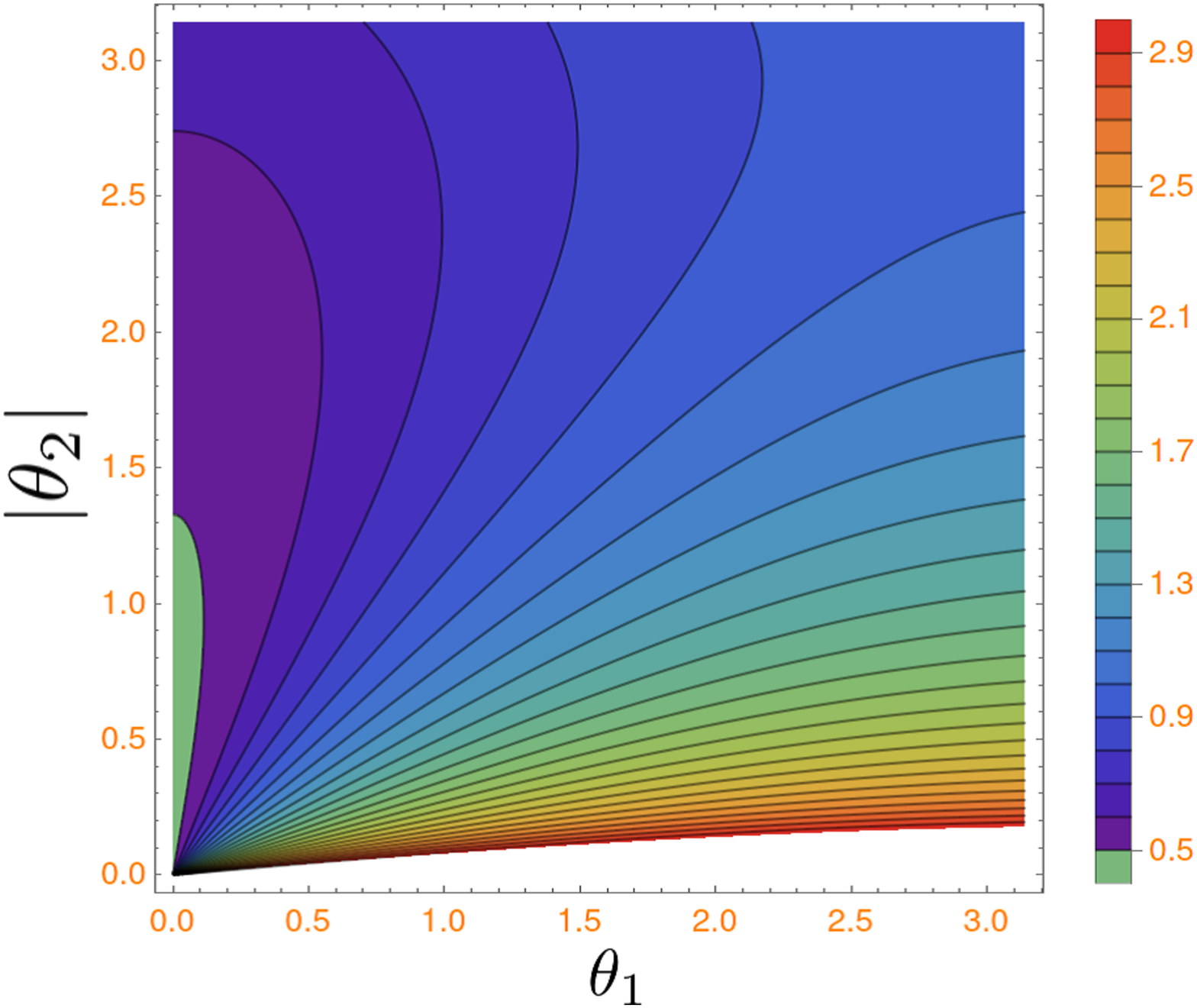}
 \caption{
Plot of $ \log_{10} (J^{-1})_{2,2}^{\infty,\frac{1}{2} } $ against $\theta_{1}$ and $\theta_{2}$ with a parameter of $q_{0}=0.33$. 
 }
 \label{fig:J_n=inf,q_0=033}
 \end{figure}
\begin{figure}[htbp]
 \includegraphics[height=3.9cm,width=7.4cm]
 {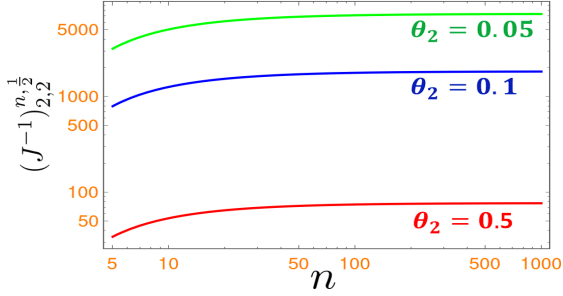}
 \caption{
 Plot of  $ (J^{-1})_{2,2}^{n,\frac{1}{2} } $ 
against $n $ with parameters of $q_{0}=0.33$,
 $\theta_{1}=2$ and $\theta_{2} = 0.5, 0.1, 0.05 $. 
 }
 \label{fig:J_theta1=2,theta2=05,01,005,001,q_0=033,n_in_5to10^4}
\end{figure}
Figs. \ref{fig:J_n=10^1,q_0=033},
\ref{fig:J_n=10^4,q_0=033}, and \ref{fig:J_n=inf,q_0=033} 
show how $\log_{10} (J^{-1})_{2,2}^{n,\frac{1}{2} }$ depends on
$\theta_{1}$ and $\theta_{2}$ with a parameter of $q_{0}=0.33$.
Since $\log_{10} (J^{-1})_{2,2}^{n,\frac{1}{2} }$ is 
an even function of $\theta_{1}$ and $\theta_{2}$, 
we set the range of the plot as $0 \le \theta_{1} , \theta_{2} \le \pi$.
Importantly, even in the limit of a large number of the participants,
$\log_{10} (J^{-1})_{2,2}^{n,\frac{1}{2} }$ is upper bounded by a finite value.
Also, for any values of $\theta _1$, $\log_{10} (J^{-1})_{2,2}^{n,\frac{1}{2} }$ does not diverge.
On the other hand, as $|\theta_{2}|$ approaches zero,
$\log_{10} (J^{-1})_{2,2}^{n,\frac{1}{2} }$ approaches infinity. 
This means that we cannot estimate the magnetic fields 
with a high precision when $|\theta_{2}|$ is small. 
Next, we consider the region where $(J^{-1})_{2,2}^{n,\frac{1}{2} }$ is rapidly increasing.
Fig. \ref{fig:J_theta1=2,theta2=05,01,005,001,q_0=033,n_in_5to10^4}
shows how $\log_{10} (J^{-1})_{2,2}^{n,\frac{1}{2} }$ depends on $n$ with a parameter of $q_{0}=0.33$,
$\theta_{1}=2$ and $\theta_{2} = 0.5, 0.1,0.05 $. 
To notice, 
the horizontal axis $n$ has the range that $5 \le n \le 10^{4} )$.
As $n$ decreases, 
$ (J^{-1})_{2,2}^{n,\frac{1}{2} }$ also decreases.
\subsection{Security proof}
We prove the tracelessness of our protocol.
In our protocol,
the measurer $M$ performs the POVM measurements as we described before.
From Eq. \eqref{eq:prob_povm}, 
the probability distribution of the POVM outcomes 
depends on $c_{i,\pm}, q_{i}$ and $\gamma^{n,m}_{ i , \pm }$.
Then, 
$c_{i,\pm}$ only depends on the form of the POVM and $q_{i}$ only depends on the initial state.
These are independent of the information of the senders,
which corresponds to the positions with non-zero magnetic fields.
Also, as explained in Appendix,
we have the following relationship.
\begin{eqnarray}
\begin{cases}
\gamma^{n,m}_{ i , \pm }
=
\sum_{ l= \max{ ( 0 , i-|R| ) } }^{\min{ (i,m) }}
\frac{
\binom{|R|}{i-l}
g_{m , \pm ,l}
}
{ 2 \binom{n}{i} }
\nonumber \\
\quad
( i =0,1,\ldots, \bigl[ \frac{n-1}{2} \bigr] )
\\
\gamma^{n,m}_{ \frac{n}{2} , \pm }
=
\sum_{ l= \max{ (0, m - \frac{n}{2} ) } }^{\min{ (\frac{n}{2} ,m) }}
\delta_{+,\pm} 
 \frac{ 
\binom{|R|}{\frac{n}{2} - l}
 g_{m,+,l}
 }
{
2
\binom{n}{ \frac{n}{2} }
}
\end{cases}
\end{eqnarray}
This means that 
$ \gamma^{n,m}_{ i , \pm } $
does not depend on the information of the sender at all.
From these results, 
we show that the probability distribution of the POVM outcomes does not depend 
on $s_{i}$ at all. 
So our protocol is traceless.\par
In the above discussion, 
we assumed that we know the information of $m$ before we perform our protocol.
Now, 
let us discuss the case that our knowledge about $m$ is actually different from the true value of $m$.
We define $m_{\rm{est}}$ as our estimation of $m$.
Since we show the tracelessness for the case of one or two senders,
we assume $m_{\rm{est}}=1$ or $m_{\rm{est}}=2$.
We choose the initial state and POVM elements depending on $m_{\rm{est}}$,
and estimate the magnetic fields based on the information of $m_{\rm{est}}$.
So the uncertainty of the estimation calculated 
above does not apply when $m_{\rm{est}}$ is less than $m$. 
Nonetheless, 
the probability distribution of the POVM outcomes is independent of the information of the sender. 
So, 
even for $m_{\rm{est}} < m$, 
the tracelessness is guaranteed in our protocol.
\section{
Conclusion
}
In conclusion,
we propose anonymous quantum sensing 
as new applications of quantum sensing network. 
We assume that there are non-zero magnetic fields at one or at most two positions. 
The purpose of this protocol is to measure the non-zero magnetic fields 
while we hide the information about which positions generate magnetic fields.
In our protocol, 
each participant is located in a different position.
First, 
entangled qubits are distributed to the participants. 
Second, 
the participants interact their own qubits with magnetic fields at their positions.
Finally, 
the qubits are sent to the measurer, 
and POVM measurements are performed by the measurer $M$. 
The advantage of our protocol is that even
if the outcomes of the POVM measurements are broadcasted 
(or are stolen by an eavesdropper),
 information of the positions with non-zero magnetic fields is still unknown for the eavesdropper in our protocol.
Since the magnetic-field sensor is used in medical science and material engineering,
our protocol could play an important role to protect confidential information 
once quantum network becomes available.\par
On the other hand, 
our protocol has a unique feature as a quantum anonymous transmission as well.
Actually, 
our protocol allows at most two senders to send messages anonymously at the same time. 
In this sense,
our protocol could also be interpreted as a quantum transmission protocol with anonymous multiple senders,
 while the conventional one allows only one sender to send messages.
\section{Acknowledgements}
This work was supported by Leading Initiative 
for Excellent Young Researchers MEXT Japan and JST presto (Grant No. JPMJPR1919) Japan.
This work was also supported by CREST (JPMJCR1774 and JPMJCR15N2 ), JST.\par
Y. Takeuchi is supported by MEXT Quantum Leap Flagship Program (MEXT Q-LEAP) Grant Number JPMXS0118067394 and JPMXS0120319794, 
and JST [Moonshot R\&D -- MILLENNIA Program] Grant Number  JPMJMS2061.
\begin{widetext}
\appendix
\section{
Fisher information and Cram\'{e}r-Rao inequality
}
\label{sec:fisher-info_and_Cramer-Rao}
We introduce Fisher information to quantify the sensitivity of the parameter estimation.
We estimate an $m$-dimensional real vector 
$ \vec{\theta} = ( \theta_{1}, \theta_{2},\ldots,\theta_{m} ) $.
The Fisher information is given as a matrix, and the $(i,j)$ component 
of the matrix is given as follows.
\begin{eqnarray}
J_{i,j}
&\equiv&
\sum_{x \in \chi }
\frac{1}{ p(x| \vec{\theta} ) }
\frac{ \partial p(x| \vec{\theta} ) }{ \partial \theta_{i} }
 \frac{ \partial p(x| \vec{\theta} ) }{ \partial \theta_{j} }
\end{eqnarray}
where $x$ denotes the outcome of the measurement, 
$\chi$ denotes all possible outcomes, and 
$p(x| \vec{\theta} )$
denotes the probability to obtain $x$ with the parameter vector $\vec{\theta}$.
By assuming the regularity of $J$, 
we obtain the following Cram\'{e}r-Rao inequality
about the variance of $\theta_{i}$.
\begin{eqnarray}
(\Delta \theta_{i})^{2} \ge \frac{1}{N} (J^{-1})_{i,i}
\end{eqnarray}
$N$ is the total number of the measurement for estimation.
Note that the minimum of estimation uncertainty 
 corresponds to the maximum of the estimation sensitivity.
 The Cram\'{e}r-Rao inequality shows that the inverse of Fisher information gives 
 the maximum of sensitivity when we perform an unbiased estimation.
 Therefore, 
we use the Fisher information to quantify the sensitivity in our paper.
\section{Definitions and Properties }
\subsection{Properties of Dicke state}
\label{sec:dicke_state}
The Dicke state $\ket{D^{n}_{k}}$ is a superposition of all combinations 
where $k$ qubits are states of $\ket{1}$ and is represented as follows.
\begin{eqnarray}
\ket{D^{n}_{k}}
&\equiv&
\frac{1}
{
\sqrt{
\binom{n}{k}
}
}
\sum_{
\substack{ x \in \{0,1 \}^{n}
\\ hw(x)=k
} } \ket{x}
\end{eqnarray}
$hw(x)$ is a hamming weight of bit string $x$.
Then, we consider the following transformation.
\begin{eqnarray}
\label{eq:complementary_of_dicke}
\ket{D^{n}_{n-k}}
&=&
\biggl(
\prod_{j \in V}
\hat{\sigma}_{x,j}
\biggr)
\ket{D^{n}_{k}}
\end{eqnarray}
\begin{eqnarray}
\braket{ D^{n}_{k} | D^{n}_{l} }
&=&
\delta_{k,l}
\end{eqnarray}
By applying $\prod_{j \in V} \hat{\sigma}_{x,j}$ on a Dicke state $\ket{D^{n}_{k}}$ 
with $k$ ( $k=0,1,\ldots, \big[ \frac{n}{2} \bigr] $), we obtain another Dicke state $\ket{D^{n}_{k}}$ 
with $k$ ( $k=n,n-1,\ldots, \big[ \frac{n}{2} \bigr] +1$ ).
The representation of the Dicke states is useful especially when the participants are divided into two sets. 
For example, 
let us consider a case when $V$ is a set of all participants 
and $S$($R$) is a set of $m$($n-m$) participants as follows.
\begin{eqnarray}
\begin{cases}
S = \{ s_{1},s_{2},\ldots, s_{m} \}
\\
R = \{ r_{1},r_{2},\ldots, r_{n-m} \}
\\
V = S \oplus R
\end{cases}
\end{eqnarray}
\begin{eqnarray}
\begin{cases}
|S| = m
\\
|R| = n-m
\\
|V| = n
\end{cases}
\end{eqnarray}
The Dicke state $\ket{D^{|V|}_{k}}$ is a superposition 
of all combinations where $k$ qubits are states of $\ket{1}$.
Note that $0\le k \le |V|=|S|+|R|$ is satisfied.
Let us consider a case that $l$ qubits in $S$ are $\ket{1}$ 
and $(k-l)$ qubits in $R$ are $\ket{1}$,
and we have the following conditions.
\begin{enumerate}
\item When $ k \le |S| \land k \le |R| $\\
\begin{eqnarray}
\begin{cases}
l=0,1,\ldots,k-1,k
\\
k-l=k,k-1,\ldots,1,0
\end{cases}
\end{eqnarray}
\item When $ k \le |S| \land k > |R| $\\
\begin{eqnarray}
\begin{cases}
l=k-|R|,k-|R|+1,\ldots,k-1,k
\\
k-l=|R|,|R|-1,\ldots,1,0
\end{cases}
\end{eqnarray}
\item When $ k > |S| \land k \le |R| $\\
\begin{eqnarray}
\begin{cases}
l=0,1,\ldots,|S|-1,|S|
\\
k-l=k,k-1,\ldots,k-|S|+1,k-|S|
\end{cases}
\end{eqnarray}
\item When $ k > |S| \land k > |R| $\\
\begin{eqnarray}
\begin{cases}
l=k-|R|,k-|R|+1,\ldots,|S|-1,|S|
\\
k-l=|R|,|R|-1,\ldots,k-|S|+1,k-|S|
\end{cases}
\end{eqnarray}
\end{enumerate}
By summarizing the above relationships, the range of $l$ is described as follows.
\begin{eqnarray}
 \max{ (0,k-|R|) } \le l \le \min{ (k,|S|) }
\end{eqnarray}
Then, we can transform $\ket{D^{|V|}_{k}}$ as follows.
\begin{eqnarray}
\label{eq:dicke_devided_two_dicke_product}
{}
&{}&
\ket{D^{|V|}_{k}}
=
\frac{1}{\sqrt{
\binom{|V|}{k}
}}
\sum_{
\substack{ x \in \{0,1 \}^{|V|}
\\ hw(x)=k
} } \ket{x}
\nonumber \\
&=&
\frac{1}{\sqrt{
\binom{|V|}{k}
}}
\sum_{ l = \max{ (0,k-|R| ) } }^{\min{ (k,|S|) }}
\biggl\{
\sum_{ \substack{
\vec{d}_{S}
=
 ( d_{S,1}, d_{S,2}, \ldots , d_{S,|S|} )
\\
\in \{ 0,1\}^{|S|} ,
hw( \vec{d}_{S} ) = l
} }
\sum_{ \substack{
\vec{e}_{R}
=
 ( e_{R,1}, e_{R,2}, \ldots , e_{R,|R|} )
\\
\in \{ 0,1\}^{|R|} ,
hw( \vec{e}_{R} ) = k-l
} }
\biggl(
\bigotimes_{j = 1}^{|S|}
\hat{\sigma}_{x,s_{j} }^{d_{S,j}}
\ket{0}_{s_{j}}
\biggr)
\biggl(
\bigotimes_{j = 1}^{|R|}
\hat{\sigma}_{x,r_{j}}^{e_{R,j}}
\ket{0}_{r_{j}}
\biggr)
\biggr\}
\nonumber \\
&=&
\frac{1}{\sqrt{
\binom{|V|}{k}
}}
\sum_{ l = \max{ ( 0 , k-|R| ) } }^{\min{ (k,|S|) }}
\biggl(
\sqrt{ \binom{|S|}{l} }
\ket{D_{l}^{|S|} }_{\otimes j \in S }
\otimes
\sqrt{ \binom{|R|}{k-l} }
\ket{D_{k-l}^{|R|} }_{\otimes j \in R }
\biggr)
\nonumber \\
\end{eqnarray}
Although Dicke state was originally represented by a single set,
we can represent the Dicke state by the sum of the direct product of two Dicke states with two sets
whose direct sum gives us the original set. By repeating this technique,
the Dicke state can be expressed by the arbitrary number of sets.
\subsection{
Bit string and Hamming weight
}
\begin{enumerate}
\item Bit-string vector \\
Let us define $r$ by using a sequence of 1-bit random numbers $r_{i} \in \{0,1\} (i=1,2,\ldots,m)$ as follows.
\begin{eqnarray}
r
&\equiv&
(r_{m}r_{m-1} \cdot r_{2}r_{1})_{(2)}
\end{eqnarray}
The row vector of this is described as follows.
\begin{eqnarray}
\vec{r} 
&\equiv&
 ( r_{1}, r_{2} , \ldots, r_{m})
 \quad
 ( \in \{0,1 \}^{m} )
\end{eqnarray}
\item Bit string with Hamming weight\\
Let us define $ f_{m,l,p} $ as the $p$-th smallest number among the $m$ bit strings 
with a hamming weight of $l(=0,1,\ldots,m)$. 
We call this a Bit string with Hamming weight.
There are $ \binom{m}{l}$ numbers of $m$ bit strings with a hamming weight of $l$. 
For example, when we have $(m,l)=(4,2)$, 
we obtain the following bit strings.
\begin{eqnarray}
f_{4,2,1}
=
 (0011)_{2}
\quad
 f_{4,2,2}
=
 (0101)_{2}
 \\
\quad
f_{4,2,3}
=
 (0110)_{2}
\quad
f_{4,2,4}
=
 (1001)_{2}
\\
f_{4,2,5}
=
 (1010)_{2}
\quad
f_{4,2,6}
=
 (1100)_{2}
\end{eqnarray}
We do not define $ f_{m,l,p} $ for $m=0$.
We define $(11\ldots1)_{2}-f_{m,l,p}$ as an operation to map an  $m$ bit string 
with a hamming weight $l$ into an $m$ bit string with a hamming weight $m-l$. 
Since we subtract $f_{m,l,p}$ (the $p$-th smallest number among the $m$ bit strings with 
a hamming weight of $l(=0,1,\ldots,m)$) from $(11\ldots1)_{2}$, 
we obtain the $p$-th largest number among $m$ bit strings with 
a hamming weight of $m-l$. 
This means that we have the $( \binom{m}{l} +1-p )$-th smallest number among $m$ bit strings 
with a hamming weight of $m-l$. 
So we obain the following.
\begin{eqnarray}
 f_{m,(m-l),( \binom{m}{l} +1 -p )}
&=&
(11\ldots1)_{2}
-
f_{m,l,p} 
\nonumber\\
\quad
( p = 1,2,\ldots \binom{m}{l} )
\end{eqnarray}
Then
\begin{eqnarray}
\label{eq:complementarity_of_hw_bit}
f_{m,l,p} + f_{m,(m-l),( \binom{m}{l} +1 -p )}
&=&
(11\ldots1)_{2}
\nonumber\\
\quad
( p = 1,2,\ldots \binom{m}{l} )
\end{eqnarray}
\item Bit-string vector with Hamming weight\\
Let us consider a vector whose components are the same 
as that of a Bit string with Hamming weight $f_{m,l,p}$ in binary notation. 
We call this a ``Bit string vector with Hamming weight''. 
For the Bit string with Hamming weight, we have 
\begin{eqnarray}
f_{m,l,p}
&=&
(a_{p,m} a_{p,m-1} \ldots a_{p,2} a_{p,1} )_{2}
\end{eqnarray}
A Bit-string vector with Hamming weight is described as follows.
\begin{eqnarray}
\vec{f}_{m,l,p}
&=&
(a_{p,1} , a_{p,2} , \ldots , a_{p,m-1} , a_{p,m} )
\end{eqnarray}
Then, from \eqref{eq:complementarity_of_hw_bit},
we obtain the following relationship.
\begin{eqnarray}
\label{eq:complementarity_of_hw_bit_vec}
\vec{f}_{m,l,p} + \vec{f}_{m,(m-l),( \binom{m}{l} +1 -p )}
&=&
(1,1,\ldots,1)
\nonumber \\
\quad
( p = 1,2,\ldots \binom{m}{l} )
\end{eqnarray}
\item Sign Vector\\
Let us define a ``sign vector'' as follows.
\begin{eqnarray}
 \vec{R} (\vec{r})
&\equiv&
 ( (-1)^{r_{1}} ,(-1)^{ r_{2}} , \ldots, (-1)^{r_{m}})
 \nonumber \\
 &\quad&
 ( \in \{-1,1 \}^{m} )
\end{eqnarray}
Let us consider the $i$-th component of the sign vector $\vec{R}$ as follows.
\begin{eqnarray}
 (\vec{R})_{i}
&=&
 (-1)^{r_{i}}
=
1-2 r_{i}
\end{eqnarray}
So we obtain the following relationship.
\begin{eqnarray}
\label{eq:relation_of_sign_vec_and_bit_vec}
 \vec{R} (\vec{r})
&=&
 (1, 1 , \ldots, 1)
-2\vec{r}
\end{eqnarray}
Then,
\begin{eqnarray}
\label{eq:complementarity_of_sign_vector}
\vec{R} ( \vec{f}_{m,l,p} )
+
\vec{R} ( \vec{f}_{m,(m-l),( \binom{m}{l} +1 -p )} )
&=&
\vec{0}
\nonumber \\
\quad
( p = 1,2,\ldots \binom{m}{l} )
\end{eqnarray}
\end{enumerate}
\subsection{
Effective phase and auxiliary coefficients
}
With $ m(=1,2,\ldots n) , l (= 0,1,\ldots m ), p ( = 1,2,\ldots \binom{m}{l} )$, 
we define an effective phase as follows.
\begin{eqnarray}
\theta_{m,l,p}
&=&
t \vec{w}
\cdot
\vec{R}( \vec{f}_{m,l,p} )
\end{eqnarray}
where $ \vec{\omega} \equiv ( \omega_{1} ,\omega_{2} , \ldots , \omega_{m} )$. 
In the main text,
we discussed the unknown parameters to be estimated, 
and $\theta_{m,l,p}$ corresponds to the parameters 
while $\vec{\omega}$ corresponds to the information to be sent.
\begin{eqnarray}
t \vec{w}
\cdot
\bigl\{
\vec{R}( \vec{f}_{m,l,p} )
+
\vec{R}( \vec{f}_{m,(m-l),( \binom{m}{l} +1 -p )} )
\bigr\}
&=&
0
\nonumber \\
\end{eqnarray}
\begin{eqnarray}
\label{eq:complementarity_of_effective_phase}
\therefore
\theta_{m,l,p}
+
\theta_{m,(m-l),( \binom{m}{l} +1 -p )}
&=&
0
\end{eqnarray}
We define coefficient $h$ as follows.
\begin{eqnarray}
h_{m,l}
&\equiv&
\sum_{ p=1}^{
\binom{m}{l}
}
e^{-
\frac{1}{2} i\theta_{m,l,p}
 }
\end{eqnarray}
where $l = 0,1,\ldots , m$.
We obtain the following.
\begin{eqnarray}
h_{ m , l}^{\ast}
 &=&
h_{ m , (m-l) }
\end{eqnarray}
Let us assume that $m$ is even and we have $l=\frac{m}{2}$. 
Then, we obtain the following.
\begin{eqnarray}
h_{ m , \frac{m}{2} }^{\ast}
 &=&
h_{ m , \frac{m}{2}}
\end{eqnarray}
where
$h_{ m , \frac{m}{2} }$ is a real number.
Let us calculate the details of this as follows.
\begin{eqnarray}
\therefore
h_{m, \frac{m}{2} }
&=&
\sum_{ p=1}^{
\binom{m}{\frac{m}{2}}
}
\cos{
\frac{ \theta_{m,\frac{m}{2},p}
}{2}
 }
\end{eqnarray}
Then, we define coefficients $g$ as follows. 
\begin{eqnarray}
g_{m,\pm,l}
&\equiv&
h_{m,l}
 \pm
h_{m,l}^{\ast}
\end{eqnarray}
where $l=0,1,\ldots,m$.
We obtain the following.
\begin{eqnarray}
\therefore
g_{m,\pm,m-l}
&=&
g_{m,\pm,l}^{\ast}
\end{eqnarray}
\begin{eqnarray}
g_{ m , k, \frac{m}{2} }
&=&
 2h_{m, \frac{m}{2} } \delta_{+,k}
 (k=\pm)
\end{eqnarray}
\begin{eqnarray}
\therefore
g_{m,k,l}
&=&
\begin{cases}
-
2
\sum_{ p=1}^{
\binom{m}{l}
}
i
\sin
\frac{ \theta_{m,l,p} }{2}
 \quad
 ( k= -)
\\
2
\sum_{ p=1}^{
\binom{m}{l}
}
\cos
\frac{ \theta_{m,l,p} }{2}
 \quad
 ( k= +)
\end{cases}
\nonumber \\
\end{eqnarray}
\section{Security}
\subsection{Tracelessness of our protocol}
\label{sec:anonymity-of-probability}
We prove that probability distribution \eqref{eq:prob_povm} is independent of 
the information of the senders $ s_{i} ( \in S) $ 
and our protocol guarantees the tracelessness define in the main text.
We consider the time evolution \eqref{eq:all_time_evolution} 
by the Hamiltonian \eqref{eq:all_hamiltonian}.
Then, 
we can consider the following transformation.
\begin{eqnarray}
\label{eq:complementary_of_all_hamiltonian}
\hat{U}^{\dagger}
 &=&
\biggl(
\prod_{j \in V}
\hat{\sigma}_{x,j}
\biggr)^{\dagger}
 \hat{U}
\biggl(
\prod_{j \in V}
\hat{\sigma}_{x,j}
\biggr)
\end{eqnarray}
From \eqref{eq:dicke_devided_two_dicke_product},
we obtain the following transformation.
\begin{eqnarray}
\braket{ D^{n}_{k} |
\hat{U}
| D^{n}_{k} }
&=&
\frac{1}
{
\binom{n}{k}
}
\sum_{ l= \max{ ( 0 , k-|R| ) } }^{\min{ (k,m) }}
{ \binom{|R|}{k-l} }
h_{m,l}
\end{eqnarray}
Since the operator $ \hat{U} $ does not contain any bit-flip operators,
the operator just induces changes in the relative phase in the Dicke state.
By using the orthonormality of the Dicke states, we obtain the following.
\begin{eqnarray}
\braket{
 D^{n}_{k}
 |
\hat{U}
|
D^{n}_{k^{\prime} }
}
&=&
\delta_{k, k^{\prime} }
\braket{
 D^{n}_{k}
 |
\hat{U}
|
D^{n}_{k }
}
\end{eqnarray}
This form is independent of the elements of the sets $S$ and $R$. 
We obtain the following.
\begin{eqnarray}
\braket{ D^{n}_{n-k} |
 \hat{U}
| D^{n}_{n-k } }
&=&
\braket{ D^{n}_{k} |
\biggl(
\prod_{j \in V}
\hat{\sigma}_{x,j}
\biggr)^{\dagger}
 \hat{U}
\biggl(
\prod_{j \in V}
\hat{\sigma}_{x,j}
\biggr)
| D^{n}_{k } }
\quad
( \because \eqref{eq:complementary_of_dicke} )
\nonumber \\
 &=&
\braket{ D^{n}_{k} |
 \hat{U}^{\dagger}
| D^{n}_{k } }
\quad
( \because \eqref{eq:complementary_of_all_hamiltonian} )
\nonumber \\
 &=&
 (
\braket{ D^{n}_{k} |
 \hat{U}
| D^{n}_{k } }
)^{\ast}
\end{eqnarray}
The absolute value of $\braket{ D^{n}_{k} |
 \hat{U}
| D^{n}_{k } } $ is equal to that of $
\braket{ D^{n}_{n-k} |
 \hat{U}
| D^{n}_{n-k } }
$. Without loss of the generality, we can assume the following range of $k$.
\begin{eqnarray}
 k
 &=&
 0,1,\ldots, \biggl[ \frac{n}{2} \biggr]
\end{eqnarray}
Let us assume that $n$ is an even number with $k= \frac{n}{2} $, and we obtain the following.
\begin{eqnarray}
 (
\braket{ D^{n}_{\frac{n}{2} } |
 \hat{U}
| D^{n}_{\frac{n}{2} } }
)^{\ast}
&=&
\braket{ D^{n}_{ \frac{n}{2} } |
 \hat{U}
| D^{n}_{ \frac{n}{2} } }
\end{eqnarray}
and
$
\braket{ D^{n}_{ \frac{n}{2} } 
|
 \hat{U}
|
D^{n}_{\frac{n}{2} } } 
$
becomes a real number.
Let us calculate the details of this as follows.
\begin{eqnarray}
\braket{ D^{n}_{ \frac{n}{2} } |
\hat{U}
| D^{n}_{ \frac{n}{2} } }
&=&
\frac{1}
{
2
\binom{n}{ \frac{n}{2} }
}
\sum_{ l= \max{ (0, m - \frac{n}{2} ) } }^{\min{ (\frac{n}{2} ,m) }}
\binom{|R|}{\frac{n}{2} - l}
 g_{m,+,l}
\end{eqnarray}
Then, 
we define the superposition of the Dicke states as follows.
\begin{eqnarray}
\ket{ \phi^{n}_{k,l} }
&=&
\begin{cases}
\frac{1}{ \sqrt{2} }
\biggl(
\ket{ D^{n}_{k} }
+l
\ket{ D^{n}_{n-k} }
\biggr)
\quad
( n \neq 2k , l = \pm )
\\
\delta_{+,l}
\ket{ D^{n}_{ \frac{n}{2} } }
( n = 2k , l = \pm )
\end{cases}
\end{eqnarray}
Then, we obtain the following relationship about
$
\braket{
\phi^{n}_{k,l} |
\hat{U}
| \phi^{n}_{ k^{\prime} , l^{\prime} }
}$.
\begin{enumerate}
\item When $n \neq 2k$\\
We use
$
\delta_{k , n-k^{\prime} }
=
\delta_{n-k , k^{\prime} }
$ and obtain the following 
\begin{eqnarray}
{}
&{}&
\braket{
\phi^{n}_{k,l} |
\hat{U}
| \phi^{n}_{ k^{\prime} , l^{\prime} }
}
\nonumber \\
&=&
\frac{1}{2}
\biggl\{
\delta_{k , k^{\prime} }
\biggl(
\bra{ D^{n}_{k} }
\hat{U}
\ket{ D^{n}_{k} }
+
l l^{\prime}
(
\bra{ D^{n}_{k} }
\hat{U}
\ket{ D^{n}_{k} }
)^{\ast}
\biggr)
+
\delta_{n-k , k^{\prime} }
\biggl(
l^{\prime}
\bra{ D^{n}_{k} }
\hat{U}
\ket{ D^{n}_{k} }
+
l
(
\bra{ D^{n}_{k} }
\hat{U}
\ket{ D^{n}_{k} }
)^{\ast}
\biggr)
\biggr\}
\nonumber \\
\end{eqnarray}
Let us calculate the details of this as follows.
\begin{eqnarray}
{}
&{}&
\braket{
\phi^{n}_{k,l} |
\hat{U}
| \phi^{n}_{ k^{\prime} , l^{\prime} }
}
\nonumber \\
&=&
\frac{1}{2}
\biggl\{
\delta_{k , k^{\prime} }
\biggl(
\bra{ D^{n}_{k} }
\hat{U}
\ket{ D^{n}_{k} }
+
l l^{\prime}
(
\bra{ D^{n}_{k} }
\hat{U}
\ket{ D^{n}_{k} }
)^{\ast}
\biggr)
+
\delta_{n-k , k^{\prime} }
\biggl(
l^{\prime}
\bra{ D^{n}_{k} }
\hat{U}
\ket{ D^{n}_{k} }
+
l
(
\bra{ D^{n}_{k} }
\hat{U}
\ket{ D^{n}_{k} }
)^{\ast}
\biggr)
\biggr\}
\nonumber \\
&=&
\frac{1}{2}
\biggl[
\delta_{k , k^{\prime} }
\biggl\{
\frac{1}
{
\binom{n}{k}
}
\sum_{ a= \max{ ( 0 , k-|R| ) } }^{\min{ (k,m) }}
\binom{|R|}{k-a}
h_{m,a}
+
l l^{\prime}
\biggl(
\frac{1}
{
\binom{n}{k}
}
\sum_{ a= \max{ ( 0 , k-|R| ) } }^{\min{ (k,m) }}
\binom{|R|}{k-a}
h_{m,a}
\biggr)^{\ast}
\biggr\}
\nonumber \\
&+&
\delta_{n-k , k^{\prime} }
\biggl\{
l^{\prime}
\frac{1}
{
\binom{n}{k}
}
\sum_{ a= \max{ ( 0 , k-|R| ) } }^{\min{ (k,m) }}
\binom{|R|}{k-a}
h_{m,a}
+
l
\biggl(
\frac{1}
{
\binom{n}{k}
}
\sum_{ a= \max{ ( 0 , k-|R| ) } }^{\min{ (k,m) }}
\binom{|R|}{k-a}
h_{m,a}
\biggr)^{\ast}
\biggr\}
\biggr]
\nonumber \\
&=&
\frac{
\delta_{k , k^{\prime} }
+
\delta_{n-k , k^{\prime} }
l^{\prime}
}
{ 2 \binom{n}{k} }
\sum_{ a= \max{ ( 0 , k-|R| ) } }^{\min{ (k,m) }}
\binom{|R|}{k-a}
g_{m , l l^{\prime} ,a}
\end{eqnarray}
Next,
We obtain the following by using
$
\delta_{k , n-k^{\prime} }
=
\delta_{n-k , k^{\prime} }
$.
\begin{eqnarray}
\braket{
 \phi^{n}_{ k^{\prime} , l^{\prime} }
|
\hat{U}
|
\phi^{n}_{k,l}
}
&=&
\braket{
\phi^{n}_{k,l}
|
\hat{U}
|
 \phi^{n}_{ k^{\prime} , l^{\prime} }
}
\end{eqnarray}
Next
\begin{eqnarray}
\braket{
\phi^{n}_{n-k,l} |
\hat{U}
| \phi^{n}_{ n-k^{\prime} , l^{\prime} }
}
&=&
(
\braket{
\phi^{n}_{k,l} |
\hat{U}
| \phi^{n}_{ k^{\prime} , l^{\prime} }
}
)^{\ast}
\end{eqnarray}
The absolute value of $\braket{
\phi^{n}_{k,l} |
\hat{U}
| \phi^{n}_{ k^{\prime} , l^{\prime} }
}$
is equal to that of
$\braket{
\phi^{n}_{n-k,l} |
\hat{U}
| \phi^{n}_{ n-k^{\prime} , l^{\prime} }
}$. Without loss of generality, we can assume
\begin{eqnarray}
 k ,k^{\prime}
 &=&
 0,1,\ldots, \biggl[ \frac{n}{2} \biggr]
\end{eqnarray}
However, we assumed $n \neq 2k$,
and so the range of $k$ is as follows.
\begin{eqnarray}
 k ,k^{\prime}
 &=&
 0,1,\ldots, \biggl[ \frac{n-1}{2} \biggr]
\end{eqnarray}
\item When $n=2k$
\begin{eqnarray}
\braket{
\phi^{n}_{\frac{n}{2},l}
|
\hat{U}
|
\phi^{n}_{ \frac{n}{2} , l^{\prime} }
}
&=&
(
\bra{ D^{n}_{ \frac{n}{2} } }
\delta_{+,l}
)
\hat{U}
(
\delta_{+,l^{\prime}}
\ket{ D^{n}_{ \frac{n}{2} } }
)
\nonumber \\
&=&
\delta_{+,l}
\delta_{+,l^{\prime}}
\braket{
 D^{n}_{ \frac{n}{2} }
 |
\hat{U}
|
 D^{n}_{ \frac{n}{2} }
 }
\end{eqnarray}
\item When the combination of $n=2k$ and $ n \neq 2k^{\prime} $\\
If $ n \neq 2k^{\prime} $, 
i.e., $ n-k^{\prime} \neq \frac{n}{2} $
\begin{eqnarray}
\braket{
\phi^{n}_{\frac{n}{2},l}
|
\hat{U}
|
\phi^{n}_{ k^{\prime} , l^{\prime} }
}
&=&
\bigl(
\bra{ D^{n}_{ \frac{n}{2} } }
\delta_{+,l}
\bigr)
\hat{U}
\biggl\{
\frac{1}{ \sqrt{2} }
\biggl(
\ket{ D^{n}_{k^{\prime}} }
+
l^{\prime}
\ket{ D^{n}_{n-k^{\prime}} }
\biggr)
\biggr\}
\nonumber \\
&=&
\frac{ \delta_{+,l} }{ \sqrt{2} }
\biggl(
\braket{
 D^{n}_{ \frac{n}{2} }
|
\hat{U}
|
 D^{n}_{k^{\prime}}
}
+
l^{\prime}
\braket{
 D^{n}_{ \frac{n}{2} }
 |
\hat{U}
|
 D^{n}_{n-k^{\prime}}
}
\biggr)
\nonumber \\
&=&
\frac{ \delta_{+,l} }{ \sqrt{2} }
\biggl(
\delta_{ \frac{n}{2} , k^{\prime} }
\braket{
D^{n}_{k^{\prime}}
|
\hat{U}
|
 D^{n}_{k^{\prime}}
}
+
l^{\prime}
\delta_{ \frac{n}{2} , n-k^{\prime} }
\braket{
D^{n}_{ n-k^{\prime} }
|
\hat{U}
|
 D^{n}_{ n-k^{\prime} }
}
\biggr)
\nonumber \\
&=&
0
\end{eqnarray}
\end{enumerate}
We use
$\braket{
\phi^{n}_{k,+ }
|
\hat{U}
|
 \phi^{n}_{ k , l }
}$ 
to calculate the probability distibution, 
and this does not depend on $S$ but depends on $m(=|S|)$.
We define as follows.
\begin{eqnarray}
\gamma^{n,m}_{ k , l }
&\equiv&
\braket{
\phi^{n}_{k,+ }
|
\hat{U}
|
 \phi^{n}_{ k , l}
}
\end{eqnarray}
Then, we consider the range of $k$ as follows.
\begin{eqnarray}
 k
 &=&
 0,1,\ldots, \biggl[ \frac{n-1}{2} \biggr]
\end{eqnarray}
We obtain
$ \delta_{n- k,k }=0$, and we have
the following transformation.
\begin{eqnarray}
\gamma^{n,m}_{ k , \pm }
&=&
\braket{
\phi^{n}_{k,+}
|
\hat{U}
|
 \phi^{n}_{ k , \pm }
}
\nonumber \\
&=&
\frac{
\delta_{k , k }
+
\delta_{n-k , k }
(\pm1)
}
{ 2 \binom{n}{k} }
\sum_{ l= \max{ ( 0 , k-|R| ) } }^{\min{ (k,m) }}
{ \binom{|R|}{k-l} }
g_{m , \pm ,l}
\nonumber \\
&=&
\frac{1}
{ 2 \binom{n}{k} }
\sum_{ l= \max{ ( 0 , k-|R| ) } }^{\min{ (k,m) }}
{ \binom{|R|}{k-l} }
g_{m , \pm ,l}
\end{eqnarray}
Then,
\begin{eqnarray}
(
\gamma^{n,m}_{ k , \pm }
)^{\ast}
&=&
\pm
\gamma^{n,m}_{ k , \pm }
\end{eqnarray}
Next, we consider the case of $ k = \frac{n}{2}$.
\begin{eqnarray}
\gamma^{n,m}_{ \frac{n}{2} , \pm }
&=&
\frac{ \delta_{+,\pm} }
{
2
\binom{n}{ \frac{n}{2} }
}
\sum_{ l= \max{ (0, m - \frac{n}{2} ) } }^{\min{ (\frac{n}{2} ,m) }}
\binom{|R|}{\frac{n}{2} - l}
 g_{m,+,l}
\nonumber \\
\end{eqnarray}
\subsection{Evaluation of uncertainty}
\label{sec:evaluation-of-uncertainty}
We explain the details of how we evaluate $(J^{-1})_{2,2}$.
Suppose that $a$ is a function of $n$ ($a=a(n)$). 
We define $\beta_{a,n}$ as follows.
\begin{eqnarray}
\beta_{a,n}
\equiv
\frac{a}{n}
\end{eqnarray}
Since we have $ |S| \le a \le \bigl[ \frac{n}{2} \bigr] $, 
we obtain the following.
\begin{eqnarray}
\frac{|S|}{n}
\le
\beta_{a,n}
\le
\frac{1}{n}
 \bigl[ \frac{n}{2} \bigr]
\end{eqnarray}
Despite the fact that the upper bound of $a$ depends on $n$,
we could consider that $a$ and $n$ are independent parameters.
In the large limit of $n$, 
we obtain $\beta_{a,\infty}$. 
We show the relationship between $a$ and $\beta_{a,\infty}$ as follows.
\begin{eqnarray}
\begin{cases}
a=O(n) \Rightarrow \beta_{a,\infty}>0
\\
a=o(n) \Rightarrow \beta_{a,\infty}=0
\end{cases}
\end{eqnarray}
About the upper bound of $ \beta_{a,n} $,
we obtain the following relationships.
\begin{eqnarray}
\begin{cases}
 \frac{1}{n} \bigl[ \frac{n}{2} \bigr]
 =
 \frac{1}{n} \cdot \frac{n}{2}
 =
 \frac{1}{2}
 \quad ( n \text{ is even} )
 \\
 \frac{1}{n} \bigl[ \frac{n}{2} \bigr]
 =
 \frac{1}{n} \cdot \frac{n-1}{2}
 =
 \frac{1}{2}
 (1 -\frac{1}{n} )
 \xrightarrow{ n \rightarrow \infty }
 \frac{1}{2}
 \quad ( n \text{ is odd} )
\end{cases}
\end{eqnarray}
Then, 
by considering $n \rightarrow \infty$,
we obtain the following.
\begin{eqnarray}
0 \le \beta_{a,\infty} \le \frac{1}{2}
\end{eqnarray}
Next, we define $l$ as follows.
\begin{eqnarray}
l
&\equiv&
\frac{2a(n-a)}{ n(n-1)}
=
\frac
{ 2 \beta_{a,n} ( 1- \beta_{a,n} )}
{ 1- n^{-1} }
\end{eqnarray}
It is worth mentioning that $(1- \frac{1}{n})$ 
is monotonically increasing function of $n$ and 
$ 2 \beta_{a,n} ( 1- \beta_{a,n} ) $ is 
monotonically increasing function at $ 
\frac{|S|}{n}
\le
\beta_{a,n}
\le
\frac{1}{n}
 \bigl[ \frac{n}{2} \bigr]
$.
Then, from $ 
\frac{|S|}{n}
\le
\beta_{a,n}
\le
\frac{1}{n}
 \bigl[ \frac{n}{2} \bigr]
$ and $5 \le n$,
the maximum value of $l$ can be achieved when
$( n , \beta_{a,n} ) = (5, \frac{1}{5} \bigl[ \frac{5}{2} \bigr] )$.
So, when $n=5$, 
we obtain the following relationships.
\begin{eqnarray}
\beta_{a,5}
&=&
\frac{1}{5} \bigl[ \frac{5}{2} \bigr]
=
\frac{2}{5}
\end{eqnarray}
\begin{eqnarray}
l
&=&
\frac
{ 2 \beta_{a,5} ( 1- \beta_{a,5} )}
{ 1- \frac{1}{5}}
=
\frac{3}{5}
\end{eqnarray}
Then, 
$l$ is minimized at $ \beta_{a,n} = \frac{|S|}{n} $
in the limit of a large $n$
\begin{eqnarray}
l
&=&
\frac
{ 2 \beta_{a,n} ( 1- \beta_{a,n} )}
{ 1- \frac{1}{n}}
=
\frac
{ 2 \frac{|S|}{n} 
 ( 1- \frac{|S|}{n} ) }
{ 1- \frac{1}{n}}
>
0
\end{eqnarray}
Here, we obtain the range of $l$ as follows.
\begin{eqnarray}
0 < l \le \frac{3}{5} 
\end{eqnarray}
When $n \rightarrow \infty$, we obtain
\begin{eqnarray}
l
&=&
 2 \beta_{a,\infty} ( 1- \beta_{a,\infty} )
\end{eqnarray}
From $0 \le \beta_{a,\infty} \le \frac{1}{2} $,
the range of $l$ at $n \rightarrow \infty$ is the following.
\begin{eqnarray}
0 \le l \le \frac{1}{2} 
\end{eqnarray}
We have the following relationship between $l$, $n$, and $\beta_{a,n} $. 
\begin{eqnarray}
\begin{cases}
\frac{ \partial l }
{ \partial \beta_{a,n} }
=
\frac
{ 2 ( 1- 2\beta_{a,n} )}
{ 1- \frac{1}{n}}
\ge
0
\\
\frac{ \partial l }
{ \partial n }
=
-
\frac
{ 2 \beta_{a,n} ( 1- \beta_{a,n} )}
{ ( n -1)^{2} }
\le
0
\end{cases}
\end{eqnarray}
By rewriting $(J^{-1})_{2,2}$ to $(J^{-1})_{2,2}^{n , \beta}$,
we obtain
the following expression.
\begin{eqnarray}
{}
&{}&
(J^{-1})_{2,2}^{n,\beta}
\nonumber \\
&=&
\frac{ 1 }
{ (1-q_{0} )
\sin^{2} { \frac{ \theta_{2} }{2} }
}
 \biggl\{
2( l^{-1} - 1 )
\biggl(
 1 - \cos { \frac{ \theta_{1} }{2} } \cos { \frac{ \theta_{2} }{2} } 
 \biggr)
+ 
\sin^{2} { \frac{ \theta_{2} }{2} } 
+
( l^{-1} - 1)^{2}
\frac{1}{ q_{0}}
\sin^{2} { \frac{ \theta_{1} }{2} }
 \biggr\}
\nonumber \\
&=&
\frac{ 1 }
{ (1-q_{0} )
\sin^{2} { \frac{ \theta_{2} }{2} }
}
 \biggl[
2 \biggl\{
\frac
{ 1- n^{-1} }
{ 2 \beta_{a,n} ( 1- \beta_{a,n} )}
- 1 \biggr\}
\biggl(
 1 - \cos { \frac{ \theta_{1} }{2} } \cos { \frac{ \theta_{2} }{2} } 
 \biggr)
 \nonumber \\
&+& 
\sin^{2} { \frac{ \theta_{2} }{2} } 
+
\biggl\{
\frac
{ 1- n^{-1} }
{ 2 \beta_{a,n} ( 1- \beta_{a,n} )}
 - 1
 \biggr\}^{2}
\frac{1}{ q_{0}}
\sin^{2} { \frac{ \theta_{1} }{2} }
 \biggr]
\end{eqnarray}
Then, we derive the following.
\begin{eqnarray}
\frac{ \partial (J^{-1})_{2,2}^{n,\beta} }
{ \partial n }
&=&
-
\frac{ 1 }
{ (1-q_{0} )
\sin^{2} { \frac{ \theta_{2} }{2} }
}
 \biggl[
2
 \biggl(
1- \cos { \frac{ \theta_{1} }{2} } \cos { \frac{ \theta_{2} }{2} }
 \biggr)
+
\frac{ 
2 ( l^{-1} - 1)
 }
{ q_{0}}
\sin^{2} { \frac{ \theta_{1} }{2} }
 \biggr]
 l^{-2}
 \frac
{ \partial l }
{ \partial n }
\ge
0
 \nonumber \\
\end{eqnarray}
\begin{eqnarray}
\frac{ \partial (J^{-1})_{2,2}^{n,\beta} }
{ \partial \beta_{a,n} }
&=&
-
\frac{ 1 }
{ (1-q_{0} )
\sin^{2} { \frac{ \theta_{2} }{2} }
}
 \biggl[
2
 \biggl(
1- \cos { \frac{ \theta_{1} }{2} } \cos { \frac{ \theta_{2} }{2} }
 \biggr)
+
\frac{ 
2 ( l^{-1} - 1)
 }
{ q_{0}}
\sin^{2} { \frac{ \theta_{1} }{2} }
 \biggr]
 l^{-2}
 \frac
{ \partial l }
{ \partial \beta_{a,n} }
\le
0
 \nonumber \\
\end{eqnarray}
From the above relationships,
$ (J^{-1})_{2,2}^{n,\beta} $ is minimized at 
$ \beta_{a,n} = \frac{1}{n} \bigl[ \frac{n}{2} \bigr] $.
More specifically, when $n$ is a odd number, 
we obtain $ \beta_{a,n} = \frac{1}{n} \bigl[ \frac{n}{2} \bigr] = \frac{1}{2} \bigl( 1- \frac{1}{n} \bigr) $,
and the minimum value is given as following.
\begin{eqnarray}
{}
&{}&
 (J^{-1})_{2,2}^{n, \frac{1}{2} \bigl( 1- \frac{1}{n} \bigr) }
\nonumber \\
&=&
\frac{ 1 }
{ (1-q_{0} )
\sin^{2} { \frac{ \theta_{2} }{2} }
}
 \biggl\{
2
\biggl(
 1 - \frac { 2}{ n+1 }
\biggr)
\biggl(
1-
\cos { \frac{ \theta_{1} }{2} } \cos { \frac{ \theta_{2} }{2} }
\biggr)
+
 \sin^{2} { \frac{ \theta_{2} }{2} } 
+
\frac{ 1}{ q_{0}}
 \biggl( 
 1 - \frac { 2}{ n+1 }
 \biggr)^{2}
\sin^{2} { \frac{ \theta_{1} }{2} }
 \biggr\}
 \nonumber \\
\end{eqnarray}
When $n$ is an even number, 
we obtain
$ \beta_{a,n} = \frac{1}{n} \bigl[ \frac{n}{2} \bigr] = \frac{1}{2} $
and the minimum is the following.
\begin{eqnarray}
 (J^{-1})_{2,2}^{n, \frac{1}{2} }
&=&
\frac{ 1 }
{ (1-q_{0} )
\sin^{2} { \frac{ \theta_{2} }{2} }
}
 \biggl\{
2 \biggl( 1- \frac{2}{n} \biggr)
\biggl(
1- \cos { \frac{ \theta_{1} }{2} } \cos { \frac{ \theta_{2} }{2} }
\biggr)
 +
 \sin^{2} { \frac{ \theta_{2} }{2} } 
+
\frac{ 1}{ q_{0}}
\biggl( 1- \frac{2}{n} \biggr)^{2} 
\sin^{2} { \frac{ \theta_{1} }{2} }
 \biggr\}
 \nonumber \\
\end{eqnarray}
Let us summarize the relationships explained above.
\begin{eqnarray}
{}
&{}&
 (J^{-1})_{2,2}^{n, \frac{1}{n} \bigl[ \frac{n}{2} \bigr] }
\nonumber \\
&=&
\frac{ 1 }
{ (1-q_{0} )
\sin^{2} { \frac{ \theta_{2} }{2} }
}
 \biggl\{
2 \biggl( 1- \frac{1} { \bigl[ \frac{n+1}{2} \bigr] } \biggr)
\biggl(
1- \cos { \frac{ \theta_{1} }{2} } \cos { \frac{ \theta_{2} }{2} }
\biggr)
 +
 \sin^{2} { \frac{ \theta_{2} }{2} } 
+
\frac{ 1}{ q_{0}}
\biggl( 1- \frac{1} { \bigl[ \frac{n+1}{2} \bigr] } \biggr)^{2} 
\sin^{2} { \frac{ \theta_{1} }{2} }
 \biggr\}
 \nonumber \\
\end{eqnarray}
By considering a limit of large $n$, 
$(J^{-1})_{2,2}^{n, \frac{1}{n} \bigl[ \frac{n}{2} \bigr] }$
 approaches to $(J^{-1})_{2,2}^{\infty,\frac{1}{2} }$.
 Therefore, we obtain the following.
\begin{eqnarray}
(J^{-1})_{2,2}^{\infty,\frac{1}{2} }
&=&
\frac{
 \sin^{2}{\frac{\theta_{1}}{2}}
+
q_{0}
(
2
-
2 \cos{\frac{\theta_{1}}{2}}\cos{\frac{\theta_{2}}{2}}
+
 \sin^{2}{\frac{\theta_{2}}{2}}
)
}
{
 (1-q_{0})q_{0}
\sin^{2}{\frac{\theta_{2}}{2}}
}
\end{eqnarray}
\end{widetext}
\bibliographystyle{apsrev4-1}
\bibliography{bibref}
\end{document}